\documentclass{article}

\usepackage{float}
\usepackage{graphicx}
\usepackage{tikz}
\usepackage{iclr2023_conference, times}
\usepackage{hyperref}

\iclrfinaltrue

\graphicspath{ {./images/} }

\title{Machine learning-assisted close-set X-ray diffraction phase identification of transition metals}
\author{Maksim Zhdanov \\
NUST MISIS \\
\texttt{dwdcodes@gmail.com} \\
\And
Andrey Zhdanov \\
Tomsk Polytechnic University \\
\texttt{zhdanov.andrei24@gmail.com} \\
}

\date{}

\begin{document}

\maketitle

\begin{abstract}
Machine learning has been applied to the problem of X-ray diffraction phase prediction with promising results. In this paper, we describe a method for using machine learning to predict crystal structure phases from X-ray diffraction data of transition metals and their oxides. We evaluate the performance of our method and compare the variety of its settings. Our results demonstrate that the proposed machine learning framework achieves competitive performance. This demonstrates the potential for machine learning to significantly impact the field of X-ray diffraction and crystal structure determination. Open-source implementation: \url{https://github.com/maxnygma/NeuralXRD}.
\end{abstract}

\section{Introduction}

The determination of crystal structure phases from X-ray diffraction data is an important task in materials science, with applications ranging from drug design \citep{datta2004crystal} to the study of complex materials for hydrogen absorption \citep{akiba2006crystal}. It's crucial to address the task of unnecessary phases of elements which can occur in the process of X-ray diffraction. Notably, oxides can occur in the process of synthesis. Traditionally, this problem has been approached through the use of techniques such as direct methods \citep{duax1972valinomycin} and density functional theory \citep{van2014validation}. However, these methods can be computationally difficult to operate with \citep{burke2012perspective} and may not always yield accurate results \citep{burke2012perspective}. In recent years, there has been a growing interest in using machine learning methods to tackle this problem.

Machine learning algorithms have the ability to learn patterns and relationships in data, making them well-suited for the analysis of large and complex datasets. They have been applied to a wide range of problems in materials science, including the prediction of material properties \citep{chibani2020machine} and the analysis of imaging data \citep{wei2019machine}. In the context of X-ray diffraction, machine learning has been used to analyze crystal structure and to classify diffraction patterns \citep{suzuki2020symmetry}.


\section{Related work}

There has been a significant amount of research in the field of using machine learning for X-ray diffraction (XRD) phase identification. Previous studies \citep{PARK2022109840, https://doi.org/10.48550/arxiv.1811.08425, https://doi.org/10.48550/arxiv.2211.08591} have applied various machine learning techniques, such as support vector machines (SVMs), artificial neural networks (ANNs), and k-nearest neighbors (k-NN), to analyze XRD patterns and identify different phases in materials. These methods have been applied to a wide range of materials, including minerals \citep{https://doi.org/10.48550/arxiv.2211.08591} and ceramics \citep{Kaufmann_Maryanovsky_Mellor_Zhu_Rosengarten_Harrington_Oses_Toher_Curtarolo_Vecchio_2020}. However, many of these studies have not focused on transition metal oxides which take an important part in many applications of material science, and there is still a need for more robust and accurate methods for XRD phase identification of these elements.

\section{Method}

Below, the method for transition metals phases identification based on machine learning and X-ray diffraction patterns is introduced. Firstly, we formulate the general setup of the approach. Then the proposed method is explained as a machine learning problem with regard to implementation details.

\subsection{Synthetic pattern generation}

Initially, the method needs a set of X-ray diffraction patterns $X=\{x_{1}, \ldots, x_{i}^{N}\}$ where $N$ is the number of materials. There are 15 materials in the presented setup consisting of transtion metals and variations of their oxides. All of the samples are obtained from he Crystallography Open Database \citep{Grazulis2009} in the format of cif files. Full list of materials is shown in Table \ref{table_1}. XRD simulations from molecular structures were obtained using Mercury software \citep{Macrae:gj5232}.

\begin{table}[H]
    \centering
    \begin{tabular}{|p{1.4in}|p{1.4in}|}
        \hline
        Basic materials & Oxides  \\ 
        \hline
        $Ti$, $V$, $Zn$, $Ni$, $Mn$, $Fe$, $Cr$, $Co$ & $Al_{2}O_{3}$, $BaO_{2}$, $CaO$, $Cu_{2}O$, $Li_{2}O$, $MgO$, $NaO_{3}$\\
        \hline  
    \end{tabular}
    \caption{Elements presented in the collected set.} 
    \label{table_1}
\end{table}

Patterns iteratively combined with each other 5 rounds to create a set of synthetic data samples $S$. Single-material instances are removed from $S$. Random shift by $-1 \leq 2\theta \leq 1$ and intensity scaling with multiplier factor $t$, $0.5 \leq t \leq 1.5$ is applied preserving order of phases.


Synthetic samples are used to conduct experiments on a more generalized sets of materials as the real-world diffraction patterns may be different in terms of $2\theta$ precision. After combining samples each pattern in $X$ is normalized in the range from 0 to 1. 

\subsection{Peak matching}

\begin{figure*}
    \centering
    \includegraphics[width=\textwidth]{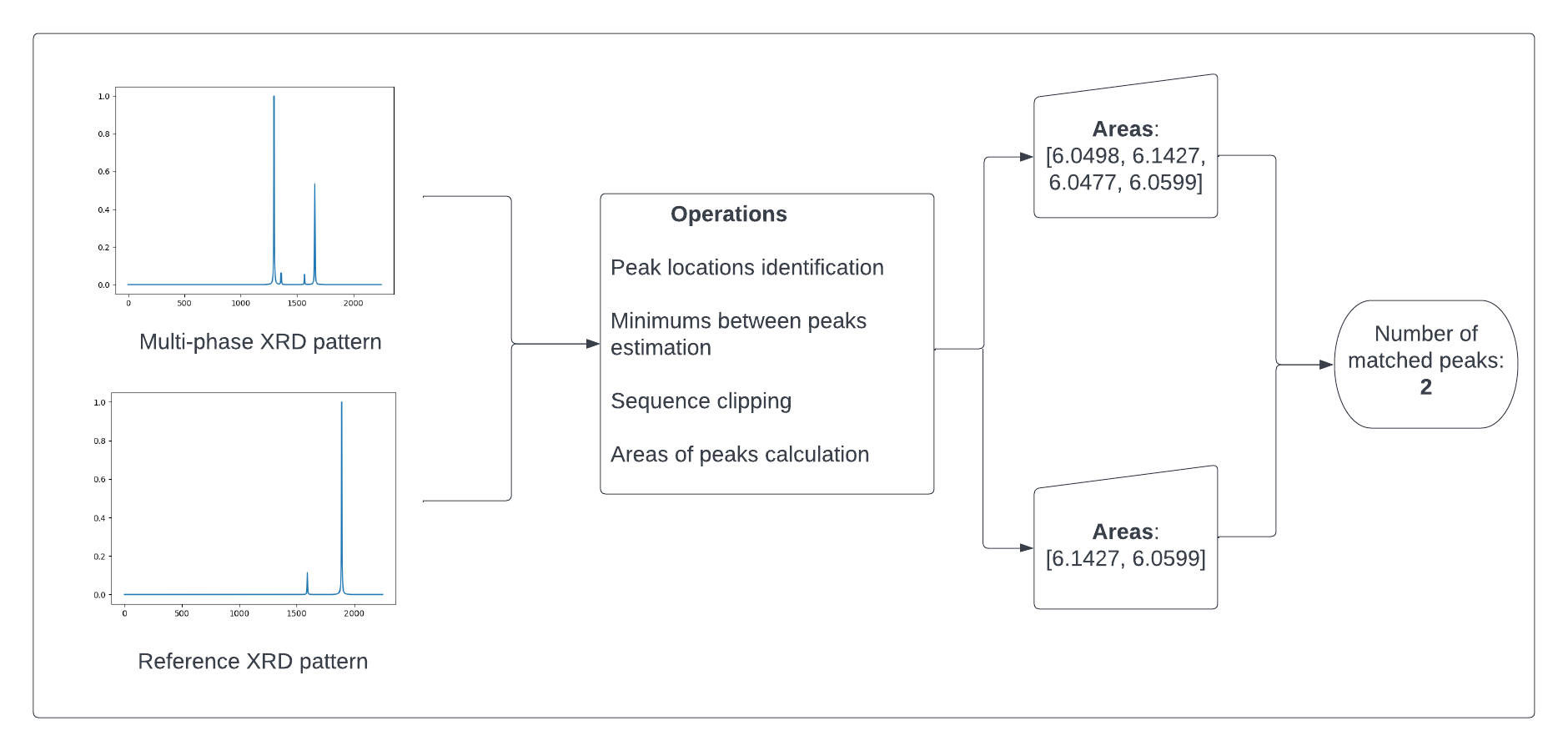}
    \caption{Peak matching algorithm overview.}
    \label{fig_2}
\end{figure*}

To identify elements presented in a sample of multi-phase diffraction patterns we propose the following peak matching algorithm. Complete overview of the approach can be seen in Figure \ref{fig_2}.
\begin{enumerate}
    \item Single sample $s_{i} \in S$ and reference sample $x_{i} \in X$ are selected.
    \item Sets of peak locations $L_{x}, L_{s}$ are computed for $x_{i}$ and $s_{i}$ by estimating local maxima $m_{i}$ and $m_{s}$ for every sample.
    \item Local minimas between each pair of peaks are calculated, then $x_{i}$ and $s_{i}$ are split on lists $P_{x}=\{p_{1}, \ldots, p_{i}\}, i=|L_{x}|$ and $P_{s}=\{p_{1}, \ldots, p_{j}\}, j=|L_{s}|$ of parts respectively.
    \item Values of $P_{x}$ and $P_{s}$ are clipped below the threshold of 0.1.
    \item Areas of $P_{x}$ and $P_{s}$ are calculated using composite trapezoidal rule. For $p_{i} \in P_{x}$ the area is obtained from 
        \begin{center}
            $\displaystyle \int_{a}^{b} p_{i}dx \approx \frac{1}{2} \sum_{j=1}^{n}(x_{j}-x_{j-1})[f(x_{j-1})+f(x_{j})]$.
        \end{center}
    Identical operation is applied for $p_{j} \in P_{s}$. Areas additionally rounded to the 4th figure after the floating point, for other approximation options see Section \ref{experiments}.
    \item Acquired areas are measured to find matching values.
\end{enumerate}

\subsection{Calibration model}


Described framework provides efficient and accurate solution to the problem. However, attempts when the algorithm fails to detect material properly because of overlapping or difficult peaks are not addressed. For this reason, we create an additional support model which uses meta-features obtained from peak matching algorithm to predict materials more precisely. 

\subsubsection{Models}

Performance analysis is conducted for particular supervised learning and ensemble models: Logistic Regression, Support Vector Machine (SVM), k-Nearest Neighbours (kNN), Decision Tree (DT), Random Forest (RF). All of the models were implemented in Scikit-Learn \citep{scikit-learn}. To evaluate performance precisely, 5-fold stratified cross-validation strategy was used based on equal distribution of samples with positive and negative matches. Features were additionally normalized by removing the mean and scaling to unit variance. Complete list of features used for training is shown below. 
\begin{itemize}
    \item The number of peaks calculated by the matching algorithm.
    \item The first index on which the condition $s_{i} > 0$ is true.
    \item The last index on which the condition $s_{i} > 0$ is true.
    \item The first index on which the condition $x_{i} > 0$ is true.
    \item The last index on which the condition $x_{i} > 0$ is true.
    \item The sum of areas of peaks for multi-phase material.
    \item Target value: is $x_{i}$ presented in $s_{i}$ or not.
\end{itemize}

\begin{table*}[t]
    \centering
    \begin{tabular}{|p{1.4in}|p{1in}|p{1in}|p{1in}|}
        \hline
        Model & Precision & Recall & F1 Score \\ [10pt]
        \hline
        No calibration model & 0.975 & 0.772 & 0.862 \\ [10pt]
        \hline
        Logistic Regression & 0.989 & 0.763 & 0.861 \\ [10pt]
        \hline
        Support Vector Machine (SVM) & \textbf{0.991} & 0.773 & 0.869 \\
        \hline
        k-Nearest Neighbours (kNN) & 0.920 & 0.834 & 0.875 \\
        \hline
        Decision Tree (DT) & 0.768 & \textbf{0.870} & 0.816 \\ [10pt]
        \hline
        Random Forest (RF), basic & 0.887 & 0.820 & 0.852 \\
        \hline
        Random Forest (RF), tuned & 0.977 & 0.813 & \textbf{0.888} \\
        \hline
    \end{tabular}
    \caption{Experiments showcase and their corresponding scores.} 
    \label{table_3}
\end{table*}



\section{Experiments} \label{experiments}

\begin{table}[H]
    \centering
    \begin{tabular}{|p{1in}|p{1in}|}
        \hline
        Rounding point & F1 Score \\ 
        \hline
        5 & 0.875 \\
        \hline
        4 & \textbf{0.888} \\
        \hline
        3 & 0.614 \\
        \hline
        2 & 0.349 \\
        \hline
    \end{tabular}
    \caption{Comparison of performance with different rounding points.} 
    \label{table_2}
\end{table}

Selected algorithms were trained on data acquired from peak matching. Standard parameters for each model are selected and compared using defined evaluation metrics. Namely, precision, recall and F1-score are used. $\displaystyle F1 = 2 * \frac{Precision * Recall}{Precision + Recall}$, is essentially the harmonic mean of precision and recall. Since the distribution of classes is highly imbalanced, F1 score addresses this problem as balances between precision and recall. Comparison of mentioned methods can be seen in Table \ref{table_3}. It's noted that some of the models did not beat the baseline without any additional supervised learning algorithm. Theoretically, this might be due to the fact that peaks positions are hard to learn since they can overlap for different pairs of materials. After careful consideration of hyperparameters, random forest with number of trees - 50, increased the maximum depth of the tree to 14 and log loss as the criterion achieves the highest F1 score among other models. Moreover, we have to consider the importance of the rounding value as matching of areas under peaks have significant dependence on it as shown in Table \ref{table_2}.

\section{Future work}
Described method has its own limitations. Close-set phase identification is still might not be accurately performing on real-world scenarios, where pattern and crystal structure may vary due to the number of reasons including but not limited to the state of diffractometer, previous conducted experiments and the flow of the current experiment. For instance, noise reduction and baseline correction techniques might have to be incorporated to handle more complex cases.

\section{Conclusion}

In conclusion, machine learning has shown to be a powerful tool for X-ray diffraction (XRD) phase identification. A variety of machine learning algorithms, such as support vector machines, random forests, and k-nearest neighbors, have been applied to analyze XRD patterns and identify different phases in materials by comparing them to a close-set database of XRD patterns. One of the key features used for this analysis is the peak area calculation, which is used to quantify the intensity of the diffraction peaks. This is an important aspect as it allows for more accurate identification of different phases in the XRD patterns. The close-set database of XRD patterns enables the identification to be performed in a more specific and targeted manner, limiting the possibility of misidentification. Presented methods have been effective in identifying phases in small datasets and a limited number of phases, but there is still a need for more robust and accurate methods for XRD phase identification on large datasets. This research highlights the potential of machine learning in XRD phase identification, especially for close-set setups, and suggests directions for future research in this field.

\bibliographystyle{iclr2023_conference}
\bibliography{ref}

\end{document}